\documentclass[aps,pra,longbibliography,twocolumn,
floatfix]{revtex4-1}
\usepackage[next]{inputenc}
\usepackage[dvips]{epsfig}
\usepackage{bbm,bm,bbold}
\usepackage{booktabs} 
\usepackage{svg} 
\usepackage{multirow}
\usepackage{hhline}
\usepackage{comment}
\usepackage{braket}
\usepackage{float}
\usepackage{stackengine}
\usepackage{amsmath,amsfonts,amssymb,amsthm, amscd, xspace, physics, enumitem, mathtools, cancel, upgreek}
\usepackage{color}
\usepackage{xcolor}
\usepackage{tikz-feynman}
\usepackage{graphicx,latexsym}

\usepackage[colorlinks=true,urlcolor=blue,citecolor=blue,linkcolor=blue]{hyperref}

\usepackage{nameref}
\usepackage{varioref}
\usepackage{cleveref}

\usepackage{natbib}

\begin{document}
\renewcommand{\vec}{\mathbf}
\renewcommand{\Re}{\mathop{\mathrm{Re}}\nolimits}
\renewcommand{\Im}{\mathop{\mathrm{Im}}\nolimits}
\newcommand\scalemath[2]{\scalebox{#1}{\mbox{\ensuremath{\displaystyle #2}}}}

\newcommand{\cmnt}[1]{\ignorespaces}

\newcommand{\jfl}[1]{{\color{blue}#1}}

\title{Exciton-polarons in monolayer semiconductors with drifting electrons}

\author{Shadman Iqbal}
\email{shadman.iqbal@monash.edu}
\affiliation{School of Physics and Astronomy, Monash University, Victoria 3800, Australia}

\author{Jesper Levinsen}
\affiliation{School of Physics and Astronomy, Monash University, Victoria 3800, Australia}

\author{Meera M. Parish}
\affiliation{School of Physics and Astronomy, Monash University, Victoria 3800, Australia}

\author{Dmitry K. Efimkin}
\email{dmitry.efimkin@monash.edu}
\affiliation{School of Physics and Astronomy, Monash University, Victoria 3800, Australia}

\begin{abstract}
Recent work has demonstrated that the optical absorption of moderately doped two-dimensional semiconductors can be described in terms of exciton-polarons. In this picture, interactions between a photoexcited exciton and a Fermi sea of excess charge carriers split the exciton spectrum into a lower-energy attractive polaron branch and a higher-energy repulsive polaron branch. Due to the negligible in-plane momentum of photons, however, optical measurements probe only exciton-polarons near zero momentum, leaving their finite-momentum physics largely unexplored. Here we propose accessing this regime by measuring the absorption in the presence of an electric bias current. In effect, this displaces the electronic Fermi sea in momentum space, such that an optically active exciton-polaron at rest acquires finite momentum relative to the charge carriers. Using a diagrammatic formulation of the exciton-polaron problem, we derive exact relations between the spectral and quasiparticle properties of current-biased exciton-polarons and their finite-momentum counterparts in an equilibrium Fermi sea. These relations show that absorption measurements with drifting electrons can reconstruct the momentum dependence of the quasiparticle properties for both branches. We estimate that experimentally accessible currents can probe a broad momentum range while remaining within the linear-response regime.
\end{abstract}

\date{\today}
\maketitle
\section{Introduction}
\label{sec:1}

Two-dimensional (2D) semiconductors have emerged as versatile platforms for exploring optoelectronic \cite{wang2012, lopezsanchez2013, mak2016, manzeli2019, shen2021, wang2022} and valleytronic phenomena \cite{schaibley2016, bussolotti2018, vitale2018, son2019, zhao2021, pu2021}. In particular, monolayer transition-metal dichalcogenides (TMDs) support tightly bound excitons that remain stable and dominate the optical response even at room temperature ~\cite{chernikov2014, hansen2024} due to their reduced dielectric screening and strong Coulomb interactions. These materials also offer an exceptional degree of tunability: their electronic and optical properties can be engineered through electrostatic doping \cite{ross2013, santos2013, chernikov2015, ye2016}, dielectric and proximity effects in van der Waals heterostructures \cite{zeng2018}, and externally applied strain \cite{shi2013, ghorbani2013, desai2014}.

Among the available tuning knobs, gate voltage is particularly powerful, providing precise and continuous control over the excitonic spectrum by adjusting the density of mobile charge carriers. As the carrier density increases, interactions between the excitons and the carriers 
lead to a shift of the excitonic resonance, observable as a peak in the absorption spectrum, while an additional redshifted peak appears. The additional resonance
was historically attributed to a trion: a weakly bound three-particle complex comprising either two electrons and a hole or two holes and an electron \cite{mak2013, ross2013, zhang2014, zhu2015, chernikov2015, cadiz2016, courtade2017}. Although this interpretation accounts for the approximate separation between the two resonances, it does not capture their systematic evolution with carrier density. This doping dependence is instead well described by exciton-polaron theory, in which a photoexcited exciton is dressed by interactions with a Fermi sea of excess charge carriers, giving rise to attractive and repulsive polaron branches that evolve continuously with doping from the trion and exciton resonances, respectively \cite{sidler2016, efimkin2017, fey2020, rana2020, efimkin2021, katsch2022, huang2023, Elaine2025,tiene2023,GalzovXT,Muir2022}.

Experimental tests of the exciton-polaron framework have so far focused predominantly on the regime of near-zero exciton momentum, directly accessible through conventional absorption and reflectance measurements. Although coupling to cavity photons permits angle-resolved spectroscopy of polaron-polaritons at finite in-plane momentum, the momenta accessible through conventional far-field optics remain small compared with the characteristic electronic momentum scales.  In principle, finite-momentum exciton-polaron physics can be probed through time-resolved photoluminescence measurements of diffusing exciton populations. However, extracting quasiparticle properties from such measurements is complicated by the finite exciton-polaron density, their mutual interactions, and disorder~\cite{kulig2018, zipfel2020, wagner2021, 
strasbourg2025, vaz2026,TransportImamoglu}. An interesting question to consider is whether the simpler, direct optical measurements may be utilised to access this finite-momentum exciton-polaron regime.
\begin{figure}[b!]
    \vspace{-0.10in}
    \centering
    \includegraphics[width=\columnwidth]{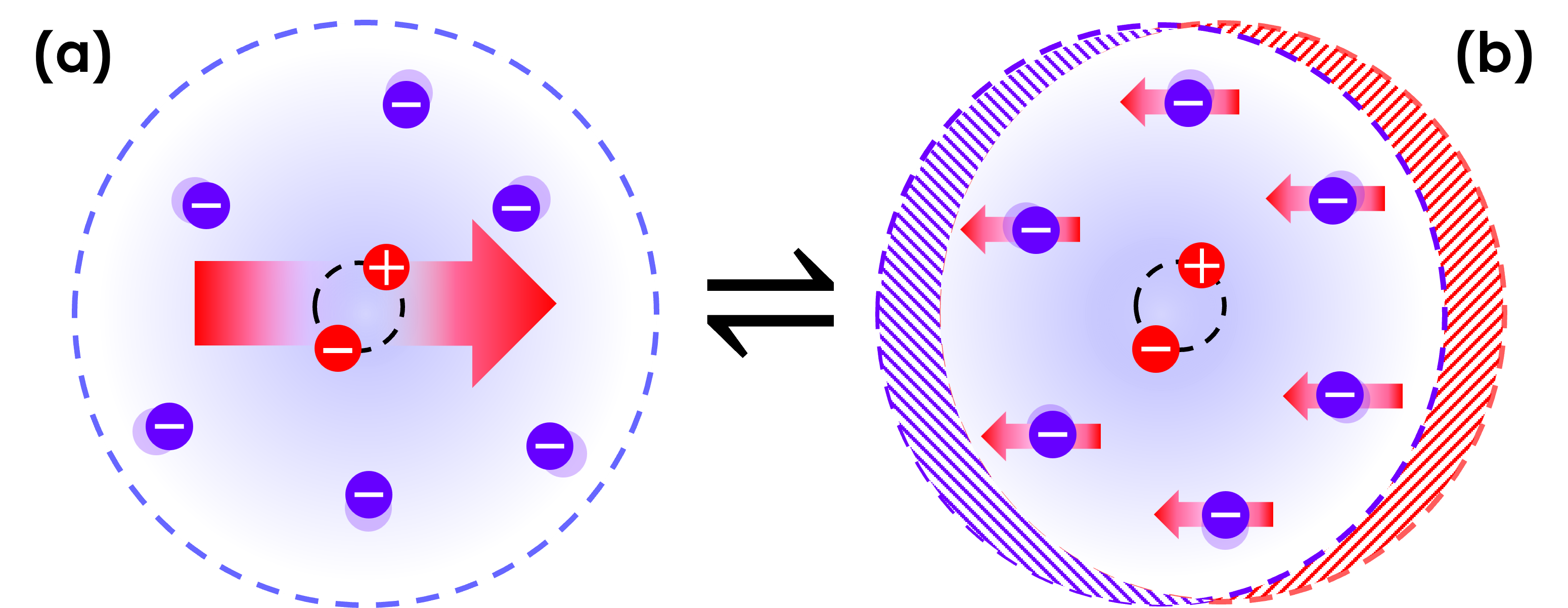}
    \caption{Schematic illustrations of (a) a finite-momentum exciton dressed by a stationary Fermi sea and (b) an exciton at rest dressed by a current-biased Fermi sea. The current-induced redistribution of electronic states, indicated by the red and violet regions, modifies Pauli blocking and the energies of Fermi-sea excitations, thereby altering the exciton's polaronic dressing.}
\label{Figure_1} 
\end{figure} 

Here, we propose accessing this regime via optical absorption measurements in the presence of an electric bias current, which displaces the electronic Fermi sea so that an optically excited exciton-polaron at rest acquires finite momentum relative to the charge carriers. This presents an intriguing parallel to the finite-momentum excitons of an unbiased system, which remain optically inaccessible, as illustrated schematically in Fig.~\ref{Figure_1}. Working within a diagrammatic exciton-polaron approach, we show that the self-energy, spectral function, and effective mass of current-biased exciton-polarons map onto those of their finite-momentum counterparts through an exact parametric scaling and frequency shift. This mapping implies that bias-dependent absorption spectra directly encode the momentum-dependent quasiparticle properties of both the attractive and repulsive polaron branches. We further estimate that, for realistic TMD mobilities, a broad range of momenta remains accessible at bias currents well within the linear-response regime. 

This paper is organized as follows: In Sec.~\ref{sec:2}, we briefly
describe the many-body $T$-matrix theory for exciton-polaron
dressing. In Sec.~\ref{sec:3}, we relate the quasiparticle
properties of finite-momentum exciton-polarons to
those of their optically accessible, current-biased zero-momentum
counterparts, and discuss the resulting momentum
dependence of their spectral features. In Sec.
\ref{sec:4}, we estimate the electric field strengths required to
access a wide range of momenta in experiment, and discuss
the limitations of our theory. 

\section{Exciton-polaron theory}
\label{sec:2}

For simplicity, we consider electron-doped MoSe$_2$ at densities low enough that only the lower spin-split conduction subband in each of the $K$ and $K'$ valleys is occupied. Circularly polarised light can selectively excite excitons in either valley. Owing to their large binding energy $\epsilon_{\mathrm{X}}\approx 250$--$500\,\mathrm{meV}$, which exceeds both the Fermi energy
$\epsilon_{\mathrm{F}}\sim 10\,\mathrm{meV}$ and the trion binding energy $\epsilon_{\mathrm{T}}\sim 30\,\mathrm{meV}$, these excitons remain well-defined mobile bosonic particles in the presence of excess charge carriers~\cite{chernikov2014,hansen2024}. Electrons populate both valleys, but those in the exciton's valley experience a short-range repulsive exchange interaction that does not favour trion formation and primarily renormalises the exciton kinetic energy~\cite{ExchangeEffects}. Exciton-polaron dressing is instead dominated by electrons in the opposite valley. 

We therefore model a valley-selected exciton coupled to the opposite-valley Fermi sea. The resulting Hamiltonian is\begin{equation}
\label{eq:hamiltonian}
\begin{split}
    \hat{H} = \sum_{\mathbf{p}}\big[\xi^{\mathrm{e}}_{\mathbf{p}} e_{\mathbf{p}}^{\dagger}e_{\mathbf{p}} \ + \xi^{\mathrm{X}}_{\mathbf{p}} X_{\mathbf{p}}^{\dagger}X_{\mathbf{p}}\big] \\ +\ \mathcal{U}\sum_{\mathbf{p},\mathbf{p^{\prime}},\mathbf{q}}e_{\mathbf{p}+\mathbf{q}}^{\dagger}X_{\mathbf{p^{\prime}}-\mathbf{q}}^{\dagger}X_{\mathbf{p^{\prime}}}e_{\mathbf{p}}.
\end{split}
\end{equation}
Here, $e_{\mathbf{p}}^{\dagger}$ ($e_{\mathbf{p}}$) and
$X_{\mathbf{p}}^{\dagger}$ ($X_{\mathbf{p}}$) denote the creation
(annihilation) operators for electrons and excitons, respectively. The first sum comprises
the single-particle kinetic energy terms, with electron and exciton
dispersions given by
$\xi^{\mathrm{e}}_{\mathbf{p}} \equiv \mathbf{p}^2/2m_{\mathrm{e}} - \epsilon_{\mathrm{F}}$
and
$\xi^{\mathrm{X}}_{\mathbf{p}} \equiv \mathbf{p}^2/2m_{\mathrm{X}} \simeq \mathbf{p}^2/4m_{\mathrm{e}}$,
respectively. Here, we focus on the case of equal electron and hole masses, $m_\mathrm{e}\approx m_\mathrm{h}$, relevant to monolayer TMDs, while the general case is treated in Appendix \ref{apx:SErelation}. Note that the exciton dispersion is measured with respect to the excitonic optical resonance in the absence of doping. Following the classical charge-dipole
interaction, the exciton-electron interaction is short-ranged
and can therefore be well approximated by a phenomenological
contact potential with coupling strength $\mathcal{U}$~\cite{efimkin2021}.\\
\begin{figure}[t!]
    \centering
    \includegraphics[width=1.0\columnwidth]{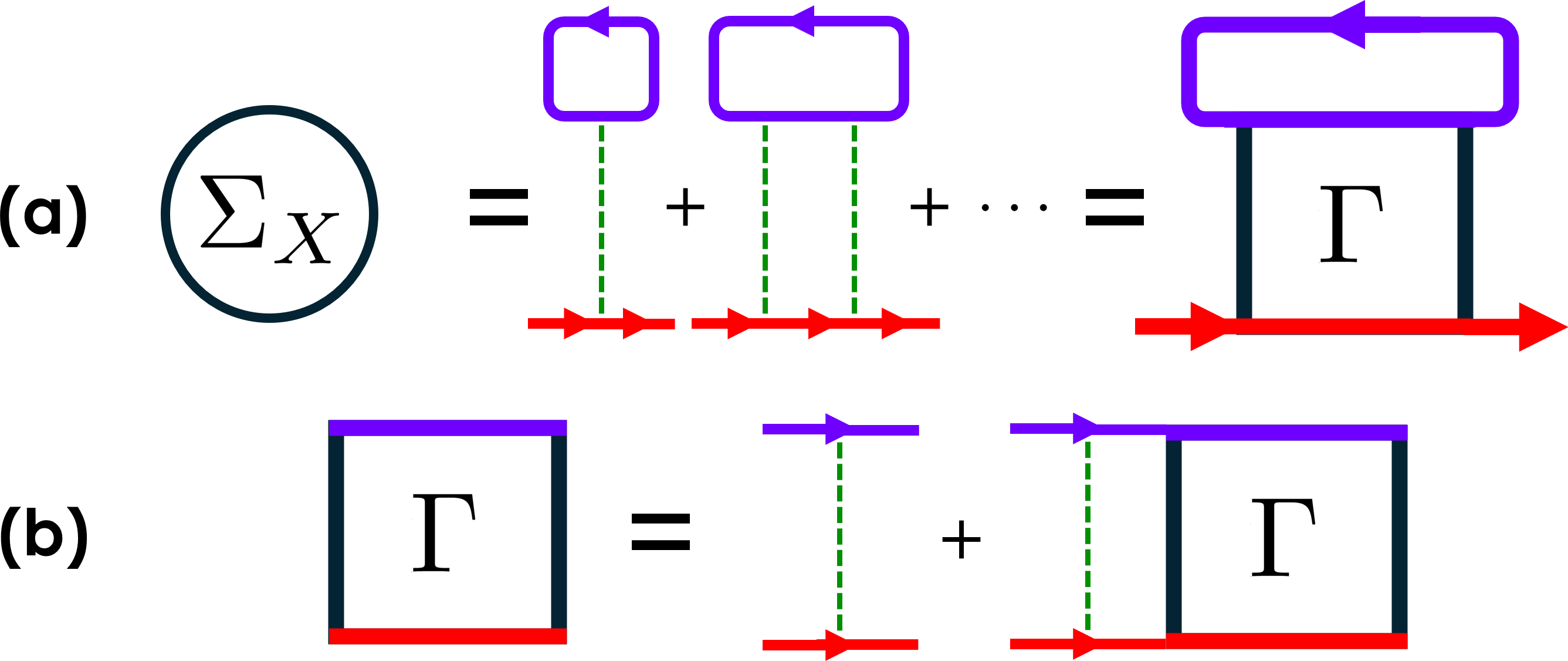}
    \caption{Diagrammatic representation of (a) the exciton self-energy $\Sigma$, expressed in terms of the vertex function $\Gamma$, and (b) the Bethe-Salpeter equation, in the ladder approximation, for the $\Gamma$-vertex describing exciton-electron scattering. Violet lines represent electrons, and red lines represent excitons.}
\label{Figure_2} 
\end{figure}

The properties of the polaron quasiparticles are encoded in the exciton spectral
function,
$\mathcal{A}_\mathrm{X}(\omega,\mathbf{p}) = -2\,\mathrm{Im}\big[G_\mathrm{X}(\omega,\mathbf{p})\big]$~\cite{sidler2016, efimkin2017, rana2020}, 
where the (retarded) exciton Green's function
$G_{\mathrm{X}}(\omega,\mathbf{p}) = \{\hbar \omega - \xi^{\mathrm{X}}_{\mathbf{p}} - \Sigma_{\mathrm{X}}(\omega,\mathbf{p})\}^{-1}$
incorporates dressing of the exciton by interactions with the
Fermi sea of excess charge carriers.
In particular, close to a polaron resonance, we have
\begin{align}
\label{eq:AXpole}
\mathcal{A}_\mathrm{X}(\omega,\mathbf{p})=\frac{2\hbar\gamma Z}{(\hbar\omega-\epsilon-\mathbf{p}^2/2m_*)^2+(\hbar\gamma)^2},
\end{align}
where $\epsilon$ is the polaron energy, $Z$ is the residue, $\gamma$ is the width, and $m_*$ is the effective mass.
Since the frequency dependence of the absorption and reflectance is
proportional to the zero-momentum limit of the exciton spectral
function, they allow the energy, residue and width to be experimentally measured, but in general not the effective mass.

The corresponding self-energy
$\Sigma_{\mathrm{X}}(\omega,\mathbf{p})$ is well described by the
ladder-diagram scattering processes sketched in
Fig.~\ref{Figure_2}(a), which collectively constitute the many-body scattering
vertex $\Gamma(\omega, \mathbf{p})$. Within this $T$-matrix approximation we have
\begin{equation}
\label{eq:selfenergy}
    \Sigma_{\mathrm{X}}(\omega,\mathbf{p}) = \sum_{\mathbf{q}}\Gamma(\hbar\omega + \xi^{\mathrm{e}}_{\mathbf{q}}, \mathbf{p} + \mathbf{q})\,n^0_{\mathrm{F}}(\mathbf{q}).
\end{equation}
Here, $n_{\mathrm{F}}^{0}(\mathbf{q})$ is the equilibrium Fermi--Dirac distribution, which we approximate by its zero-temperature step-function form, i.e., $n_{\mathrm{F}}^{0}(\mathbf{q})=\Theta(-\xi^{\mathrm{e}}_{\mathbf{q}})$. The vertex function $\Gamma(\omega,\mathbf{q})$ itself
satisfies the Bethe-Salpeter integral equation in the ladder
approximation (shown diagrammatically in
Fig.~\ref{Figure_2}(b)). As the interaction constant $\mathcal{U}$ does not depend on
frequency or momentum, this simplifies to an algebraic equation,
$\Gamma = \mathcal{U} +\mathcal{U}K\Gamma$,
whose kernel is calculated as
\begin{equation}
\label{eq:kernel}
    \mathcal{K}(\omega, \mathbf{p}) = \sum_{\mathbf{q}} \cfrac{1-n^0_{\mathrm{F}}(\mathbf{q})}{\hbar \omega - \xi^\mathrm{X}_{\vec{p}-\vec{q}} - \xi^\mathrm{e}_{\vec{q}} +i \delta},
\end{equation}
where $\delta$ denotes the radiative broadening of the exciton resonance in the absence of electrons. The kernel in Eq.~\eqref{eq:kernel}
depends solely on $\mathbf{p}$ and $\omega$. After integration, the vertex function $\Gamma(\omega,\mathbf{p})$ can be conveniently presented as 
\begin{equation}
\label{eq:manybodyvertex}
       \Gamma(\omega,\mathbf{p}) = \frac{3\pi \hbar ^2}{m_{\mathrm{e}}}\frac{1}{\log \big[\epsilon_\mathrm{T} / \Omega(\omega, \mathbf{p})\big] + i\pi},
\end{equation}
where $\epsilon_\mathrm{T} = \Lambda \exp[-3\pi\hbar^2/(\mathcal{U}m_{\mathrm{e}})]$ is the binding energy of an exciton-electron bound state serving as a simplified model of a trion, and $\Lambda$ is the ultraviolet energy cutoff required by the short-range potential approximation. This relation allows us to express $\Gamma(\omega,\mathbf{p})$ in terms of the optically measurable $\epsilon_\mathrm{T}$, eliminating both $\mathcal{U}$ and $\Lambda$. We can write the energy in Eq.~(\ref{eq:manybodyvertex}) as
\begin{equation}
\begin{split}
\label{eq:energy}
     \Omega =& \frac{1}{2}\Bigg(\hbar \omega - \frac{\mathbf{p}^2}{12m_{\mathrm{e}}} - \frac{p_{\mathrm{F}}^2}{4m_{\mathrm{e}}} + \\ & s\sqrt{\Big\{\hbar \omega - \frac{(p+p_{\mathrm{F}})^2}{4m_{\mathrm{e}}}\ \Big\}\Big\{\hbar \omega - \frac{(p-p_{\mathrm{F}})^2}{4m_{\mathrm{e}}}\ \Big\}}\ \Bigg),
\end{split}
\end{equation}
where $s = \text{sgn} (\hbar \omega - \mathbf{p}^2/4m_{\mathrm{e}} - \epsilon_{\mathrm{F}}/2)$
and $p_{\mathrm{F}}$ is the Fermi momentum. 

Equations~(\ref{eq:selfenergy})--(\ref{eq:energy}) form the backbone of the exciton-polaron theory, and they enable us, in principle, to evaluate the momentum dependence of the spectral function for an exciton with finite centre-of-mass momentum $\mathbf{p}$ dressed by a Fermi sea at rest: a regime that has thus far remained experimentally elusive. We now demonstrate that the same finite-momentum physics is exactly encoded in absorption measurements performed instead under a DC current bias.

\section{Current-biased polarons}
\label{sec:3}
We turn to the complementary configuration sketched in
Fig.~\ref{Figure_1}(b), where a DC electric bias sets the Fermi sea
into motion while the optically active exciton remains at rest. The bias current-induced steady state can be modelled as a linear
momentum shift of the equilibrium Fermi-Dirac distribution, i.e., $n_{\mathrm{F}}(\mathbf{q}) \simeq n^0_{\mathrm{F}}(\mathbf{q}+\mathbf{q}_0)$. This
follows from the Boltzmann equation in the relaxation-time
approximation \cite{mahan}: under a static field $\mathbf{E}$, the steady-state
distribution satisfies
\begin{equation}
\label{eq:boltzmann}
-\frac{e\mathbf{E}}{\hbar}\cdot\frac{\partial n_{\mathrm{F}}(\mathbf{q})}{\partial \mathbf{q}} = -\frac{n_{\mathrm{F}}(\mathbf{q})-n^0_{\mathrm{F}}(\mathbf{q})}{\tau},
\end{equation}
with $\tau$ denoting a momentum-independent relaxation time. Solving Eq.~(\ref{eq:boltzmann}) to first order in $\mathbf{E}$ gives
\begin{equation}
n_{\mathrm{F}}(\mathbf{q}) \simeq n^0_{\mathrm{F}}(\mathbf{q}) + \frac{e\tau \mathbf{E}}{\hbar} \cdot\frac{\partial n^0_{\mathrm{F}}(\mathbf{q})}{\partial \mathbf{q}}.
\end{equation}
To the same order, this is equivalent to a rigid shift of the
equilibrium distribution, $n_{\mathrm{F}}(\mathbf{q}) \simeq n^0_{\mathrm{F}}(\mathbf{q}+\mathbf{q}_0)$,
with $\mathbf{q}_0 = e\tau\mathbf{E}/\hbar \simeq \mu m_{\mathrm{e}} \mathbf{E}$, where $\mu = e\tau/\hbar m_{\mathrm{e}}$ is the mobility. Here we have identified the transport mass as the band mass since the bands of monolayer TMDs are nearly parabolic near $K$ and $K'$.
\begin{figure}[t]
    \centering
    \includegraphics[width=1.0\columnwidth]{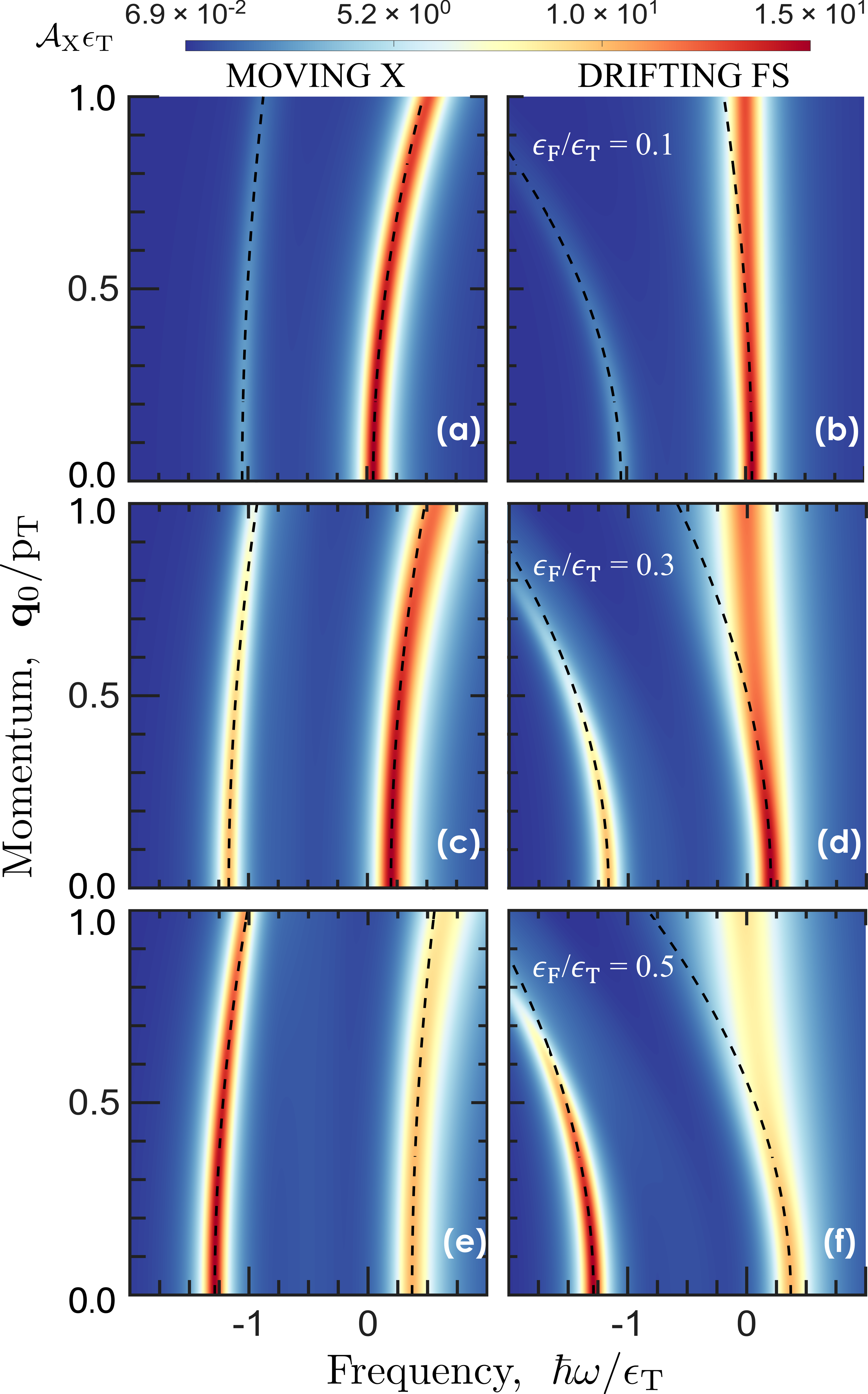}
    \caption{Exciton spectral function at various dopings, as a function
of momentum $q_0$ and frequency $\omega$. The left column [(a), (c), and (e)]
shows the spectral function $\mathcal{A}_\mathrm{X}(\omega,\mathbf{p}=\mathbf{q}_0)$ for a moving exciton (X) in an
equilibrium Fermi sea; the right column [(b), (d), and (f)] shows the
current-biased spectral function $\mathcal{A}^\mathrm{D}_\mathrm{X}(\omega,\mathbf{q}_0)$ of an exciton at rest in a Fermi sea (FS) drifting
with momentum $\mathbf{q}_0$. Both columns are shown at dopings
$\epsilon_{\mathrm{F}}/\epsilon_{\mathrm{T}}=0.1,\ 0.3,\ 0.5$ (top to
bottom). Dashed lines show effective-mass fits to the quasiparticle
peak positions.}
\label{Figure_3} 
\end{figure}

In general, the resulting spectral function
$\mathcal{A}^\mathrm{D}_\mathrm{X}(\omega,\mathbf{p},\mathbf{q}_0)$
depends on both the exciton's own momentum $\mathbf{p}$ and the
Fermi-sea (FS) drift momentum $\mathbf{q}_0$. We restrict ourselves to
optically active excitons with $\mathbf{p}=0$ and introduce the
compact notation
$\mathcal{A}^\mathrm{D}_\mathrm{X}(\omega,\mathbf{q}_0) \equiv
\mathcal{A}^\mathrm{D}_\mathrm{X}(\omega,\mathbf{0},\mathbf{q}_0)$.
A detailed derivation, presented in
Appendix~\ref{apx:SErelation}, shows that the spectral function of an exciton-polaron in motion with an equilibrium Fermi sea is connected with $\mathcal{A}^\mathrm{D}_\mathrm{X}(\omega,\mathbf{q}_0)$ as follows
\begin{equation}
\label{eq:specrel}
\mathcal{A}_\mathrm{X}(\hbar \omega,\mathbf{p})=\mathcal{A}^\mathrm{D}_\mathrm{X}\!\left(\hbar \omega-\frac{\mathbf{p}^2}{4m_{\mathrm{e}}}, \frac{\mathbf{p}}{2}\right),
\end{equation}
involving only a frequency shift and a momentum rescaling. This relation 
represents the central result of this paper. Yet, as shown in Fig.~\ref{Figure_3}, the momentum and
frequency dependence of their spectral functions, plotted for
different doping levels, appear markedly different, and this
connection is far from obvious in the spectra themselves.

In both pictures, the exciton spectral function exhibits two resonances corresponding to the attractive (AP) and repulsive (RP) exciton-polaron branches. As anticipated, in the limit of vanishing doping and momentum, their energies approach
$\epsilon_{\mathrm{AP}}\rightarrow -\epsilon_{\mathrm{T}}$
and
$\epsilon_{\mathrm{RP}}\rightarrow 0$,
so that the energy splitting between them approaches the trion binding energy $\epsilon_{\mathrm{T}}$.
\begin{figure}[t!]
    \centering
    \includegraphics[width=\columnwidth]{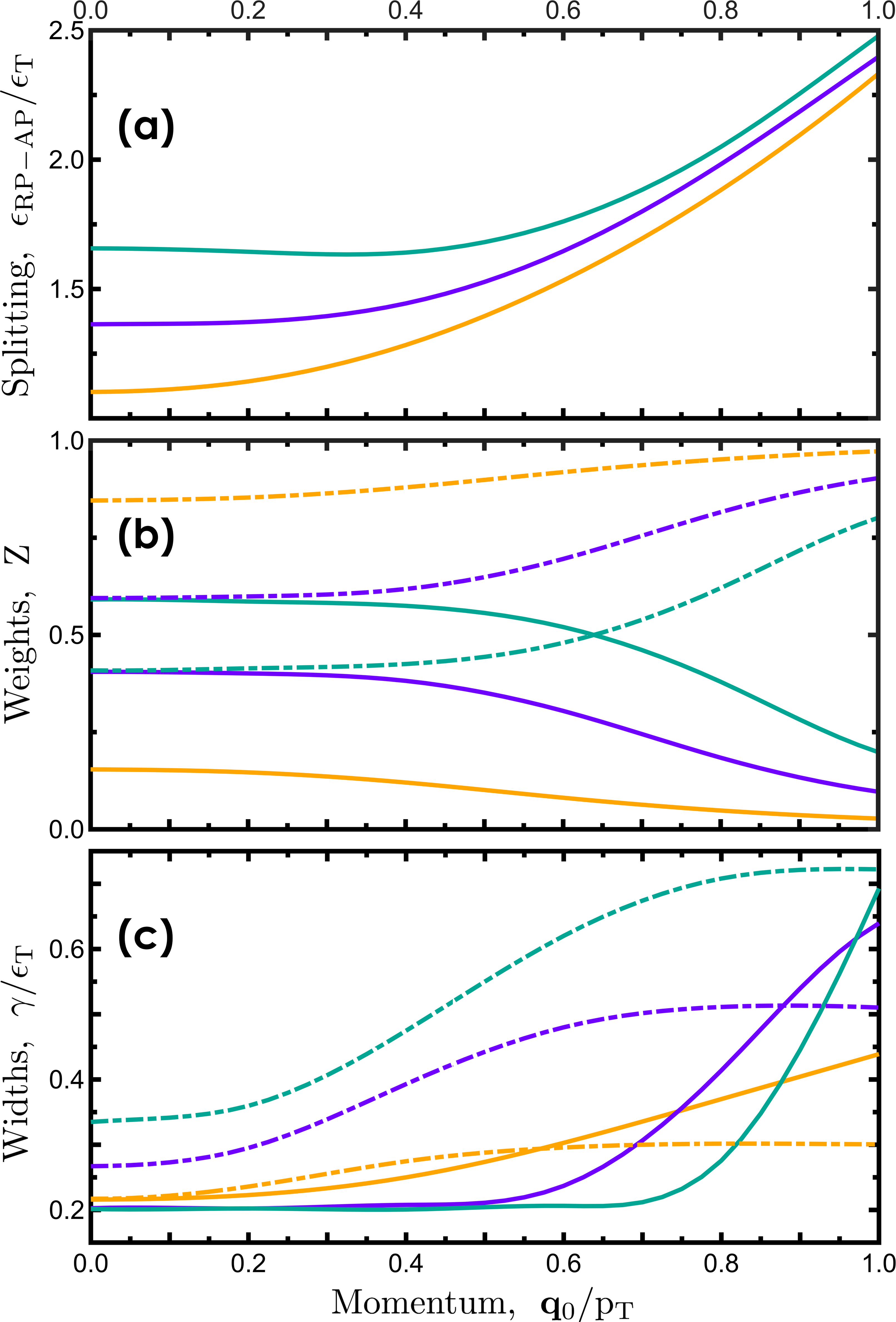}
    \caption{Momentum dependence of the splitting between RP and AP peak positions (a), peak widths (b) and spectral weights (c) for polarons in the current-biased TMD monolayer. We define the reference momentum scale $p_{\mathrm{T}}=\sqrt{2m_{\mathrm{e}}\epsilon_{\mathrm{T}}}$. Orange, violet and teal lines correspond to doping levels $\epsilon_{\mathrm{F}}/\epsilon_\mathrm{T}\ =$ 0.1, 0.3 and 0.5, respectively. In (b) and (c), solid lines correspond to the AP branch while dashed lines correspond to the RP branch.}
\label{Figure_4} 
\end{figure}
The current-biased and moving-exciton pictures nevertheless exhibit distinct branch dispersions. The current-biased AP branch has markedly greater curvature than its moving-exciton counterpart throughout the plotted range of $\mathbf{q}_0$, while the difference for the RP branch is most pronounced at low momenta. The frequency shift $\mathbf{p}^2/4m_{\mathrm{e}}$ connecting the two pictures is essential to this behaviour. This common shift, however, leaves the peak separation, spectral weights $Z$, and the broadenings $\gamma$ unchanged.

The corresponding peak separations, spectral weights, and broadenings in the moving-exciton picture can also be obtained  directly from Fig.~\ref{Figure_4} via the simple momentum rescaling $\mathbf{p}=2\mathbf{q}_0$. These results are qualitatively consistent with previous theoretical studies of finite-momentum Fermi polarons in 3D and (quasi-)2D 
cold-atom systems \cite{Massignan2011, schmidt2012,ngampruetikorn2012,parish2021, hu2022b,milczewski2024,HuiHu}. 


\begin{figure}[t!]
    \label{fig:mass pic1}
    \centering
    \includegraphics[width=\columnwidth]{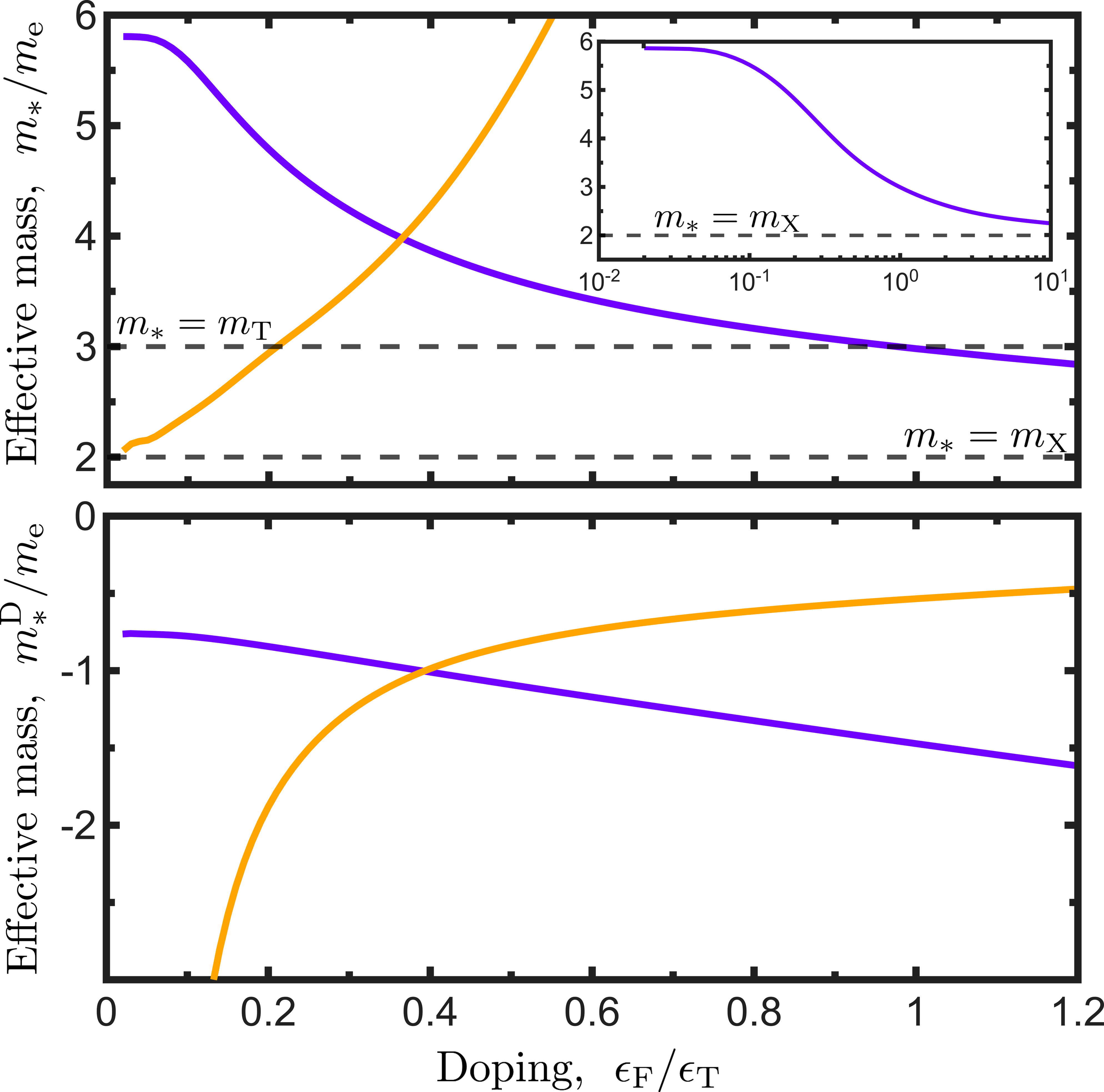}
    \caption{Doping dependence of the AP (violet) and RP (orange) polaron effective masses for the finite-momentum exciton picture in (a). Dashed lines mark the trion mass $m_{\mathrm{T}}=3m_{\mathrm{e}}$ and the bare exciton mass $m_{\mathrm{X}}=2m_{\mathrm{e}}$. \textbf{Inset:} Although the AP branch recovers the trion energy at $\epsilon_\mathrm{F}\to 0$, its effective mass shows no trace of $m_{\mathrm{T}}$, approaching $m_{\mathrm{X}}$ instead in the finite-momentum exciton picture as doping increases. Doping dependence of the AP (violet) and RP (orange) polaron effective masses for the drifting Fermi sea picture has been depicted in (b).}
\label{Figure_5} 
\end{figure}

The polaron dispersions are well described by the effective-mass approximation over a broad momentum range that, for the AP branch, extends to the depletion region, as shown by the parabolic dashed curves in Fig.~\ref{Figure_3}. The effective masses $m_*$ and their current-biased counterparts $m^\mathrm{D}_*$ differ between the two pictures but are not independent. Equation~(\ref{eq:specrel}) relates the poles of the respective dressed Green's functions, yielding the following relation for both the AP and RP branches:
\begin{equation}
\label{eq:massrel}
\frac{m^\mathrm{D}_*}{m_{\mathrm{e}}} = \frac{1}{4(m_{\mathrm{e}}/m_*) - 2}.
\end{equation}

The doping dependence of the AP and RP effective masses is shown in Figs.~\ref{Figure_5}(a) and (b) for the moving-exciton and drifting-FS pictures, respectively. In the moving-exciton picture, the AP mass does not approach the trion mass in the limit of vanishing doping, as one might naively expect (similar results have been obtained in the context of ultracold atomic gases~\cite{schmidt2012,ngampruetikorn2012}). Instead, it remains substantially larger, at approximately $6m_{\mathrm{e}}$. Over a wider doping range, the AP mass approaches the bare exciton mass $m_\mathrm{X}$, as shown in the inset of Fig.~\ref{Figure_5}, with no apparent relation to the trion mass. By contrast, the RP mass approaches $m_\mathrm{X}$ at low doping and diverges as doping increases, alongside the pronounced broadening of the RP branch noted above. 

In the drifting-FS picture, the fitted masses $m_*^\mathrm{D}$ are
negative for both exciton-polaron branches throughout the doping range considered
[Fig.~\ref{Figure_5}(b)]. This sign follows directly from
Eq.~(\ref{eq:massrel}), which can be written as
$m_*^\mathrm{D}=m_{\mathrm{e}} m_*/[2(2m_{\mathrm{e}}-m_*)]$. As shown in
Fig.~\ref{Figure_5}(a), the moving-exciton masses satisfy
$m_*>m_\mathrm{X}=2m_{\mathrm{e}}$ at finite doping, so the denominator is
negative. Furthermore, the RP mass $m_*^\mathrm{D}$ diverges as doping vanishes
because its moving-exciton counterpart approaches $2m_{\mathrm{e}}$. The negative fitted masses need not be interpreted as inertial masses. With the exciton at rest, they characterize how electron drift modifies
the absorption spectrum. They are nevertheless experimentally accessible:
$m_*^\mathrm{D}$ can be extracted from drift-dependent spectra and
converted to the physical effective masses of moving exciton-polarons, whose
finite-momentum dispersion is otherwise optically inaccessible. 

\section{Conclusions and outlook}
\label{sec:4}
The drifting Fermi sea envisaged in this work is well within reach of current experiments. To estimate the range of momenta available, 
we adopt an electron mobility of $\mu \approx 2{,}000~\mathrm{cm^2/V s}$, well below the Hall mobilities reaching $22{,}000~\mathrm{cm^2/V s}$ reported in monolayer MoSe$_2$ at $T=1.8~\mathrm{K}$~\cite{liu2025}. Assuming the linear drift relation $q_0=\mu m_{\mathrm{e}} \abs{\mathbf{E}}$, an electron effective mass $m_{\mathrm{e}}\approx0.8m_0$, where $m_0$ is the free-electron mass, and a field strength $\abs{\mathbf{E}}\lesssim2\times10^3~\mathrm{V/cm}$ yields a momentum shift $q_0/p_{\mathrm{T}}\lesssim 0.30$--$0.38$, where $p_{\mathrm{T}}=\sqrt{2m_{\mathrm{e}}\epsilon_{\mathrm{T}}}$ and $\epsilon_{\mathrm{T}}\approx25$--$40~\mathrm{meV}$. This is sufficient to resolve the momentum dependence of the AP and RP effective masses identified in Sec.~\ref{sec:3}, well within the regime where the effective-mass approximation holds and the AP branch retains most of its spectral weight. It is furthermore at field strengths comfortably below those where nonlinear transport and self-heating effects have been reported in monolayer MoS\textsubscript{2}~\cite{smithe2018}.

Finite-momentum impurities in a stationary Fermi sea have been investigated experimentally 
in the context of ultracold atomic gases~\cite{koschorreck2012,scazza2017,hennebichler2026}, 
with particular attention to their momentum-dependent spectra and polaron effective mass. 
Furthermore, there has been recent related work exploring moving impurities in a quasi-2D Bose-Einstein condensate~\cite{henke2026}. 
Our proposal offers a complementary solid-state approach, in which the relative impurity-medium motion is instead generated by driving the Fermi sea.

An interesting extension of the present work would be to explore exciton-polaron physics in crossed electric and magnetic fields. A perpendicular magnetic field quantizes the electronic spectrum into macroscopically degenerate Landau levels and splits the repulsive-polaron branch into discrete exciton--cyclotron resonances~\cite{efimkin2018,smolenski2019, liu2020, wang2020}. These features are expected to be sensitive to an additional in-plane electric field, which lifts the Landau-level degeneracy and induces a transverse drift of the electronic states. Crossed fields could therefore provide electrical control over magneto-polaron resonances and offer optical access to finite-momentum exciton-polaron physics in the quantum Hall regime.

To conclude, we have demonstrated that absorption spectroscopy under current bias reveals exciton-polaron properties over a broad momentum range beyond the reach of conventional optical measurements. Our estimates for monolayer TMDs establish the feasibility of this approach at realistic mobilities and electric field strengths.

\section*{Acknowledgements}
We acknowledge fruitful discussions with Zheng Sun. Furthermore, J. L. and D. K. E. acknowledge support from the ARC Discovery Project No. DP240100569, and M.~M.~P. acknowledges support from ARC Future Fellowship FT200100619. The authors acknowledge
the use of the large language model Claude (Anthropic)~\cite{anthropic2026claude} in the preparation of the manuscript and symbolic
verification of analytical results. The authors conceived the
project and carried out the research.
\appendix
\section{Self-energy relation} \label{apx:SErelation}
In this Appendix, we derive Eq.~\eqref{eq:specrel}, which relates the
spectral function of a moving exciton-polaron in an equilibrium Fermi sea
to that of an exciton-polaron at rest in a drifting Fermi sea. The
derivation applies to an arbitrary mass ratio, whereas the main text
focuses on $m_{\mathrm{X}}/m_{\mathrm{e}}=2$, appropriate for TMD monolayers.

We begin with the vertex function $\Gamma$ defined in
Eq.~\eqref{eq:manybodyvertex}. For arbitrary masses, it takes the form
\begin{equation}
\label{eq:generalgammavertex}
       \Gamma(\omega,\mathbf{p}) = \frac{2\pi \hbar ^2}{\mu}\frac{1}{\log \big[\frac{\epsilon_\mathrm{T}}{\Omega(\omega, \mathbf{p})}\big] + i\pi},
\end{equation}
where $\mu=m_{\mathrm{X}}m_{\mathrm{e}}/(m_{\mathrm{X}}+m_{\mathrm{e}})$
is the reduced mass of the exciton--electron pair. The energy entering
the logarithm is
\begin{equation}
\begin{split}
\label{eq:energySM}
     \Omega =& \frac{1}{2}\Bigg(\hbar \omega + \frac{ \mu \mathbf{p}^2}{m_\mathrm{X}^2} - \frac{\mathbf{p}^2}{2m_\mathrm{X}} - \frac{m_\mathrm{e} \epsilon_{\mathrm{F}}}{m_\mathrm{X}} + \\ & s\sqrt{\Big\{\hbar \omega - \frac{m_\mathrm{e} \epsilon_{\mathrm{F}}}{m_\mathrm{X}} - \frac{\mathbf{p}^2}{2m_\mathrm{X}}\ \Big\}^2 - \frac{p^2_{\mathrm{F}}\mathbf{p}^2}{m_\mathrm{X}^2}}\ \Bigg),
\end{split}
\end{equation}
where $s = \text{sgn} (\hbar \omega + \epsilon_{\mathrm{F}} - p_\mathrm{F}^2/2\mu - \mathbf{p}^2/2m_\mathrm{X})$. To relate the two pictures, we express the Bethe--Salpeter kernel
in Eq.~\eqref{eq:kernel} as
\begin{align}
\label{eq:vertexrelation}
\begin{split}
    & \mathcal{K}_{\mathrm{D}}(\omega, \mathbf{Q},\mathbf{q}_0)= \sum_{\boldsymbol{\kappa}} \frac{1 -n^0_{\mathrm{F}}(m_\mathrm{e} \mathbf{Q}/M + \boldsymbol{\kappa} - \mathbf{q}_0)}{\hbar \omega - \frac{\mathbf{Q}^2}{2M} -\frac{\boldsymbol{\kappa}^2}{2\mu} + \epsilon_{\mathrm{F}}}\\
    &= \sum_{\boldsymbol{\kappa}} \frac{1 - n^0_{\mathrm{F}}(m_\mathrm{e} \widetilde{\mathbf{Q}}/M + \boldsymbol{\kappa})}{\hbar \widetilde{\omega} - \frac{\widetilde{\mathbf{Q}}^2}{2M} - \frac{\boldsymbol{\kappa}^2}{2\mu} + \epsilon_{\mathrm{F}}}= \mathcal{K}(\widetilde{\omega}, \widetilde{\mathbf{Q}}) 
\end{split}
\end{align}
Here, $M=m_{\mathrm{X}}+m_{\mathrm{e}}$ is the total mass of the
exciton--carrier pair, $\mathbf{Q}$ is its total momentum, as illustrated
in Fig.~\ref{Figure_2}(b), and $\boldsymbol{\kappa}$ is its relative
momentum. We have introduced the transformed momentum $\widetilde{\mathbf{Q}} \equiv\ \mathbf{Q}-M\mathbf{q}_0/m_\mathrm{e}$ and frequency $\widetilde{\omega} \equiv\ \omega + \{\widetilde{\mathbf{Q}}^2 - \mathbf{Q}^2\}/2M$. The kernel relation immediately gives the corresponding relation
between the vertex functions: $\Gamma_{\mathrm{D}}(\omega, \mathbf{Q},\mathbf{q}_0)=\Gamma(\widetilde{\omega}, \widetilde{\mathbf{Q}})$. For an exciton at rest in the drifting Fermi sea,
Eq.~\eqref{eq:selfenergy} then gives
\begin{align}
\label{eq:selfenergy2}
\begin{split}
    \Sigma^{\mathrm{D}}_{\mathrm{X}}(\omega, 0,  \mathbf{q}_0) = \sum_{\boldsymbol{\kappa}} \Gamma^{\mathrm{D}}\big(\hbar \omega + \xi^{\mathrm{e}}_{\boldsymbol{\kappa}},\boldsymbol{\kappa},\mathbf{q}_0\big)n^0_{\mathrm{F}}(\boldsymbol{\kappa} - \mathbf{q}_0)& \\
    = \sum_{\boldsymbol{\kappa}} \Gamma\big(\hbar \omega +     \xi^{\mathrm{e}}_{\boldsymbol{\kappa}} + \frac{m_{\mathrm{X}}\mathbf{q}_0^2}{2m^2_{\mathrm{e}}},\boldsymbol{\kappa} - \frac{m_{\mathrm{X}}\mathbf{q}_0}{m_{\mathrm{e}}}\big)n^0_{\mathrm{F}}(\boldsymbol{\kappa}).&
\end{split}
\end{align} 
It is instructive to compare it with the self-energy for a moving exciton in a stationary Fermi sea
\begin{equation}
\label{eq:selfenergy1}
    \Sigma_{\mathrm{X}}(\omega,\mathbf{p}) = \sum_{\boldsymbol{\kappa}}\Gamma(\hbar \omega + \xi^{\mathrm{e}}_{\boldsymbol{\kappa}}, \boldsymbol{\kappa} - \mathbf{p})n^0_{\mathrm{F}}(\boldsymbol{\kappa})
\end{equation}
Here, we have used angular symmetry to replace $\abs{\boldsymbol{\kappa}+\mathbf{p}}$ with $\abs{\boldsymbol{\kappa}-\mathbf{p}}$ without changing the result. The two self-energies are thus related by a momentum rescaling and an energy shift:
\begin{align}
\label{eq:sigmarel2}
\Sigma^{\mathrm{D}}_{\mathrm{X}}(\hbar \omega, \mathbf{q}_0) = \Sigma_{\mathrm{X}}(\hbar \omega+ \frac{m_{\mathrm{X}}\mathbf{q}_0^2}{2m^2_{\mathrm{e}}}, \frac{m_{\mathrm{X}}\mathbf{q}_0}{m_{\mathrm{e}}}). 
\end{align}

The spectral function of an exciton at rest in the drifting Fermi sea
is therefore
\begin{align}
\begin{split}
&\mathcal{A}^\mathrm{D}_{X}(\omega,\mathbf{q}_0) = -2\Im\Big[\frac{1}{\hbar \omega - \Sigma^\mathrm{D}_{\mathrm{X}}(\omega,\mathbf{0},\mathbf{q}_0)}\Big]\\
&=-2\Im\Big[\frac{1}{\hbar \omega - \Sigma_{\mathrm{X}}(\hbar \omega+ \frac{m_{\mathrm{X}}\mathbf{q}_0^2}{2m^2_{\mathrm{e}}}, \frac{m_{\mathrm{X}}\mathbf{q}_0}{m_{\mathrm{e}}})}\Big]
\end{split}    
\end{align}
whereas that of a moving exciton in a stationary Fermi sea is 
\begin{equation}
\mathcal{A}_{X}(\omega,\mathbf{p}) = -2\Im\Big[\frac{1}{\hbar   \omega - \frac{\mathbf{p}^2}{2m_{\mathrm{X}}} - \Sigma_{\mathrm{X}}(\omega,\mathbf{p})}\Big] 
\end{equation} 
Comparing these expressions gives the spectral-function mapping
in both directions: 
\begin{equation}
\begin{split}
&\mathcal{A}^\mathrm{D}_{X}(\hbar \omega,\mathbf{q}_0)=\mathcal{A}_{X}\left(\hbar \omega +\frac{m_{\mathrm{X}}\mathbf{q}_0^2}{2m^2_{\mathrm{e}}}, \frac{m_{\mathrm{X}}\mathbf{q}_0}{m_{\mathrm{e}}}\right) \\
&\mathcal{A}_{X}(\hbar \omega,\mathbf{p}) = \mathcal{A}^\mathrm{D}_{X}\left(\hbar \omega -\frac{\mathbf{p}^2}{2m_{\mathrm{X}}},\frac{m_{\mathrm{e}}\mathbf{p}}{m_{\mathrm{X}}}\right)
\end{split}
\end{equation}

Close to a pole, we can expand the driven spectral function similarly to Eq.~\eqref{eq:AXpole}:
\begin{align}
\label{eq:AXDpole}
\mathcal{A}_\mathrm{X}^\mathrm{D}(\omega,\mathbf{p})=\frac{2\hbar\gamma Z}{(\hbar\omega-\epsilon-p^2/2m_*^\mathrm{D})^2+(\hbar\gamma)^2}.
\end{align}
%
This allows us to directly relate the effective masses of the AP and RP branches
in the two pictures
\begin{equation}
\label{eq:massrelgeneral}
\frac{m^\mathrm{D}_*}{m_{\mathrm{e}}} = \left[\frac{m_{\mathrm{X}}^2}{m_{\mathrm{e}}^2}\left( \frac{m_{\mathrm{e}}}{m_{*}} \right)-\frac{m_{\mathrm{X}}}{m_{\mathrm{e}}}\right]^{-1}.
\end{equation}
Setting $m_{\mathrm{e}} = m_{\mathrm{e}}$ and $m_{\mathrm{X}} = 2m_{\mathrm{e}}$ we recover Eqs.~(\ref{eq:specrel}) and ~(\ref{eq:massrel}) from the main text of the paper.   

\bibliography{XP_Biased}

@article{Elaine2025,
  title = {Valley-Mediated Singlet- and Triplet-Polaron Interactions and Quantum Dynamics in a Doped ${\mathrm{WSe}}_{2}$ Monolayer},
  author = {Ni, Yue and Huang, Di and Liang, Danfu and Liu, Albert and Liu, Xiaohui and Sampson, Kevin and Liu, Zhida and Quan, Jianmin and Watanabe, Kenji and Taniguchi, Takashi and Efimkin, Dmitry K. and Levinsen, Jesper and Parish, Meera M. and Li, Xiaoqin},
  journal = {Phys. Rev. Lett.},
  volume = {134},
  issue = {3},
  pages = {036901},
  numpages = {6},
  year = {2025},
  month = {Jan},
  publisher = {American Physical Society},
  doi = {10.1103/PhysRevLett.134.036901},
  url = {https://link.aps.org/doi/10.1103/PhysRevLett.134.036901}
}

@Article{Muir2022,
author={Muir, Jack B.
and Levinsen, Jesper
and Earl, Stuart K.
and Conway, Mitchell A.
and Cole, Jared H.
and Wurdack, Matthias
and Mishra, Rishabh
and Ing, David J.
and Estrecho, Eliezer
and Lu, Yuerui
and Efimkin, Dmitry K.
and Tollerud, Jonathan O.
and Ostrovskaya, Elena A.
and Parish, Meera M.
and Davis, Jeffrey A.},
title={Interactions between Fermi polarons in monolayer WS2},
journal={Nature Communications},
year={2022},
month={Oct},
day={18},
volume={13},
number={1},
pages={6164},
issn={2041-1723},
doi={10.1038/s41467-022-33811-x},
url={https://doi.org/10.1038/s41467-022-33811-x}
}

@article{HuiHu,
  title = {Attractive Fermi polarons at nonzero temperatures with a finite impurity concentration},
  author = {Hu, Hui and Mulkerin, Brendan C. and Wang, Jia and Liu, Xia-Ji},
  journal = {Phys. Rev. A},
  volume = {98},
  issue = {1},
  pages = {013626},
  numpages = {9},
  year = {2018},
  month = {Jul},
  publisher = {American Physical Society},
  doi = {10.1103/PhysRevA.98.013626},
  url = {https://link.aps.org/doi/10.1103/PhysRevA.98.013626}
}

@article{GalzovXT,
	Author = {M. M. Glazov},
	Doi = {10.1063/5.0012475},
	Journal = {J. Chem. Phys.},
	Pages = {034703},
	Title = {Optical properties of charged excitons in two-dimensional semiconductors},
	Url = {https://aip.scitation.org/doi/10.1063/5.0012475},
	Volume = {153},
	Year = {2020}}

@article{TransportImamoglu,
  title = {Transport of Neutral Optical Excitations Using Electric Fields},
  author = {Cotle{\ifmmode \mbox{\c{t}}\else \c{t}\fi{}}, Ovidiu and Pientka, Falko and Schmidt, Richard and Zarand, Gergely and Demler, Eugene and Imamoglu, Atac},
  journal = {Phys. Rev. X},
  volume = {9},
  issue = {4},
  pages = {041019},
  numpages = {33},
  year = {2019},
  month = {Oct},
  publisher = {American Physical Society},
  doi = {10.1103/PhysRevX.9.041019},
  url = {https://link.aps.org/doi/10.1103/PhysRevX.9.041019}
}

@book{alexandrov2012,
  title={{Advances in Polaron Physics}},
  author={Alexandrov, A. S. and Devreese, J. T.},
  isbn={9783642261015},
  url={https://link.springer.com/book/10.1007/978-3-642-01896-1},
  year={2012},
  publisher={Springer Berlin, Heidelberg},
  doi = {https://doi.org/10.1007/978-3-642-01896-1}
}

@book{baym1991,
  title={{Landau Fermi-Liquid Theory: Concepts and Applications}},
  author={Baym, G. and Pethick, C.},
  isbn={9783527617166},
  url={https://books.google.com.au/books?id=xmiV4YSEjE4C},
  year={1991},
  publisher={John Wiley \& Sons, Ltd},
  doi = {https://doi.org/10.1002/9783527617159}
}

@article{bussolotti2018,
doi = {10.1088/2399-1984/aac9d7},
year = {2018},
month = {jun},
publisher = {IOP Publishing},
volume = {2},
number = {3},
pages = {032001},
author = {Bussolotti, F. and Kawai, H. and Ooi, Z. E. and Chellappan, V. and Thian, D. and Pang, A. L. C. and Goh, K. E. J.},
title = {{Roadmap on finding chiral valleys: screening 2D materials for valleytronics}},
journal = {Nano Futures}
}

@article{cadiz2016,
    author = {Cadiz, F. and Tricard, S. and Gay, M. and Lagarde, D. and Wang, G. and Robert, C. and Renucci, P. and Urbaszek, B. and Marie, X.},
    title = {{Well separated trion and neutral excitons on superacid treated ${\mathrm{MoS}}_{2}$ monolayers}},
    journal = {Appl. Phys. Lett.},
    volume = {108},
    number = {25},
    pages = {251106},
    year = {2016},
    month = {06},
    doi = {10.1063/1.4954837}
}

@article{chernikov2014,
  title = {{Exciton Binding Energy and Nonhydrogenic Rydberg Series in Monolayer ${\mathrm{WS}}_{2}$}},
  author = {Chernikov, A. and Berkelbach, T. C. and Hill, H. M. and Rigosi, A. and Li, Y. and Aslan, B. and Reichman, D. R. and Hybertsen, M. S. and Heinz, T. F.},
  journal = {Phys. Rev. Lett.},
  volume = {113},
  issue = {7},
  pages = {076802},
  numpages = {5},
  year = {2014},
  month = {Aug},
  publisher = {American Physical Society},
  doi = {10.1103/PhysRevLett.113.076802}
}

@article{chernikov2015,
  title = {{Electrical Tuning of Exciton Binding Energies in Monolayer ${\mathrm{WS}}_{2}$}},
  author = {Chernikov, A. and van der Zande, A. M. and Hill, H. M. and Rigosi, A. F. and Velauthapillai, A. and Hone, J. and Heinz, T. F.},
  journal = {Phys. Rev. Lett.},
  volume = {115},
  issue = {12},
  pages = {126802},
  numpages = {6},
  year = {2015},
  month = {Sep},
  publisher = {American Physical Society},
  doi = {10.1103/PhysRevLett.115.126802}
}

@article{Massignan2011,
	author = {Massignan, P. and Bruun, G. M.},
	date = {2011/11/01},
	doi = {10.1140/epjd/e2011-20084-5},
	isbn = {1434-6079},
	journal = {The European Physical Journal D},
	number = {1},
	pages = {83--89},
	title = {Repulsive polarons and itinerant ferromagnetism in strongly polarized Fermi gases},
	url = {https://doi.org/10.1140/epjd/e2011-20084-5},
	volume = {65},
	year = {2011}}

@article{courtade2017,
  title = {{Charged excitons in monolayer ${\mathrm{WSe}}_{2}$: Experiment and theory}},
  author = {Courtade, E. and Semina, M. and Manca, M. and Glazov, M. M. and Robert, C. and Cadiz, F. and Wang, G. and Taniguchi, T. and Watanabe, K. and Pierre, M. and Escoffier, W. and Ivchenko, E. L. and Renucci, P. and Marie, X. and Amand, T. and Urbaszek, B.},
  journal = {Phys. Rev. B},
  volume = {96},
  issue = {8},
  pages = {085302},
  numpages = {12},
  year = {2017},
  month = {Aug},
  publisher = {American Physical Society},
  doi = {10.1103/PhysRevB.96.085302}
}

@article{desai2014,
    author = {Desai, S.
B. and Seol, G. and Kang, J. S. and Fang, H. and Battaglia, C. and Kapadia, R. and Ager, J. W. and Guo, J. and Javey, A.},
    title = {{Strain-Induced Indirect to Direct Bandgap Transition
in Multilayer ${\mathrm{WSe}}_{2}$}},
    journal = {Nano Lett.},
    volume = {14},
    number = {8},
    pages = {4592-4597},
    year = {2014},
    month = {07},
    doi = {10.1021/nl501638a}
}

@article{efimkin2017,
  title = {Many-body theory of trion absorption features in two-dimensional semiconductors},
  author = {Efimkin, D. K. and MacDonald, A. H.},
  journal = {Phys. Rev. B},
  volume = {95},
  issue = {3},
  pages = {035417},
  numpages = {10},
  year = {2017},
  month = {Jan},
  publisher = {American Physical Society},
  doi = {10.1103/PhysRevB.95.035417}
}

@article{efimkin2018,
  title = {Exciton-polarons in doped semiconductors in a strong magnetic field},
  author = {Efimkin, D. K. and MacDonald, A. H.},
  journal = {Phys. Rev. B},
  volume = {97},
  issue = {23},
  pages = {235432},
  numpages = {12},
  year = {2018},
  month = {Jun},
  publisher = {American Physical Society},
  doi = {10.1103/PhysRevB.97.235432}
}

@article{efimkin2021,
  title = {{Electron-Exciton Interactions in the Exciton-Polaron Problem}},
  author = {Efimkin, D. K. and Laird, E. K. and Levinsen, J. and Parish, M. M. and MacDonald, A. H.},
  journal = {Phys. Rev. B},
  volume = {103},
  number = {7},
  year = {2021},
  doi = {10.1103/PhysRevB.103.075417}
}

@article{fey2020,
  title = {{Theory of exciton-electron scattering in atomically thin semiconductors}},
  author = {Fey, C. and Schmelcher, P. and Imamoglu, A. and Schmidt, R.},
  journal = {Phys. Rev. B},
  volume = {101},
  issue = {19},
  pages = {195417},
  numpages = {13},
  year = {2020},
  month = {May},
  publisher = {American Physical Society},
  doi = {10.1103/PhysRevB.101.195417}
}

@article{frohlich1954,
author = { H.   Fr\"{o}hlich },
title = {Electrons in lattice fields},
doi = {10.1080/00018735400101213},
journal = {Adv. Phys.},
volume = {3},
number = {11},
pages = {325-361},
year  = {1954},

}

@article{frohlich1950,
author = { H.   Fr\"{o}hlich and H.   Pelzer  and  S.   Zienau },
title = {Properties of slow electrons in polar materials},
doi = {10.1080/14786445008521794},
journal = {Philos. Mag. J. Sci.},
volume = {41},
number = {314},
pages = {221-242},
year  = {1950},
publisher = {Taylor & Francis},


}

@article{ghorbani2013,
  title = {{Strain-dependent modulation of conductivity in single-layer transition-metal dichalcogenides}},
  author = {Ghorbani-Asl, M. and Borini, S. and Kuc, A. and Heine, T.},
  journal = {Phys. Rev. B},
  volume = {87},
  issue = {23},
  pages = {235434},
  numpages = {6},
  year = {2013},
  month = {Jun},
  publisher = {American Physical Society},
  doi = {10.1103/PhysRevB.87.235434}
}

@article{hansen2024,
author = {Hansen, K. R. and Colton, J. S. and Whittaker-Brooks, L.},
title = {{Measuring the Exciton Binding Energy: Learning from a Decade of Measurements on Halide Perovskites and Transition Metal Dichalcogenides}},
journal = {Adv. Opt. Mat.},
volume = {12},
number = {3},
pages = {2301659},
doi = {10.1002/adom.202301659},
year = {2024}
}

@article{henke2026,
  title = {Realization of repulsive polarons in the strongly correlated regime},
  author = {Henke, R. and Levinsen, J. and Parish, M. M. and Boronat, J. and Astrakharchik, G. E. and Moritz, H. and Cabrera, C. R.},
  journal = {Phys. Rev. X},
  pages = {},
  year = {2026},
  month = {Aug},
  publisher = {American Physical Society},
  doi = {10.1103/m5qj-j8bm}
}

@misc{hennebichler2026,
      title={{The moving Fermi polaron}}, 
      author={Hennebichler, J. and Erlenstedt, R. and Dobler, E. and Baroni, C. and Grimm, R. and Caldara, M. and Bruun, G. and Massignan, P.},
      year={2026},
      eprint={2606.21567},
      archivePrefix={arXiv},
      url={https://arxiv.org/abs/2606.21567}, 
}

@article{holstein1959,
title = {{Studies of polaron motion: Part II. The ``small" polaron}},
journal = {Ann. Phys.},
volume = {8},
number = {3},
pages = {343-389},
year = {1959},
issn = {0003-4916},
doi = {https://doi.org/10.1016/0003-4916(59)90003-X},
author = {T. Holstein},
}

@article{huang2023,
  title = {{Quantum Dynamics of Attractive and Repulsive Polarons in a Doped ${\mathrm{MoSe}}_{2}$ Monolayer}},
  author = {Huang, D. and Sampson, K. and Ni, Y. and Liu, Z. and Liang, D. and Watanabe, K. and Taniguchi, T. and Li, H. and Martin, E. and Levinsen, J. and Parish, M. M. and Tutuc, E. and Efimkin, D. K. and Li, X.},
  journal = {Phys. Rev. X},
  volume = {13},
  issue = {1},
  pages = {011029},
  numpages = {8},
  year = {2023},
  month = {Mar},
  publisher = {American Physical Society},
  doi = {10.1103/PhysRevX.13.011029}
}

@article{hu2022b,
  title = {{Raman spectroscopy of Fermi polarons}},
  author = {Hu, H. and Liu, X.-J.},
  journal = {Phys. Rev. A},
  volume = {106},
  issue = {6},
  pages = {063306},
  numpages = {7},
  year = {2022},
  month = {Dec},
  publisher = {American Physical Society},
  doi = {10.1103/PhysRevA.106.063306}
}

@article{katsch2022,
  title = {{Excitonic theory of doping-dependent optical response in atomically thin semiconductors}},
  author = {Katsch, F. and Knorr, A.},
  journal = {Phys. Rev. B},
  volume = {105},
  issue = {4},
  pages = {045301},
  numpages = {18},
  year = {2022},
  month = {Jan},
  publisher = {American Physical Society},
  doi = {10.1103/PhysRevB.105.045301}
}

@article{koschorreck2012,
author={Koschorreck, M.
and Pertot, D.
and Vogt, E.
and Fr{\"o}hlich, B.
and Feld, M.
and K{\"o}hl, M.},
title={{Attractive and repulsive Fermi polarons in two dimensions}},
journal={Nature},
year={2012},
day={01},
volume={485},
number={7400},
pages={619-622},
doi={10.1038/nature11151}
}

@article{kulig2018,
  title = {{Exciton Diffusion and Halo Effects in Monolayer Semiconductors}},
  author = {Kulig, M. and Zipfel, J. and Nagler, P. and Blanter, S. and Sch\"uller, C. and Korn, T. and Paradiso, N. and Glazov, M. M. and Chernikov, A.},
  journal = {Phys. Rev. Lett.},
  volume = {120},
  issue = {20},
  pages = {207401},
  numpages = {6},
  year = {2018},
  month = {May},
  publisher = {American Physical Society},
  doi = {10.1103/PhysRevLett.120.207401}
}

@article{landaupekar1948,
    author = "Landau, L. D. and Pekar, S. I.",
    title = {{The Effective Mass of the Polaron}},
    doi = "10.1016/b978-0-08-010586-4.50072-9",
    journal = "J. Exp. Theor. Phys.",
    volume = "18",
    pages = {478-483},
    year = "1948"
}

@article{liu2020,
  title = {{Landau-Quantized Excitonic Absorption and Luminescence in a Monolayer Valley Semiconductor}},
  author = {Liu, E. and van Baren, J. and Taniguchi, T. and Watanabe, K. and Chang, Y.-C. and Lui, C. H.},
  journal = {Phys. Rev. Lett.},
  volume = {124},
  issue = {9},
  pages = {097401},
  numpages = {8},
  year = {2020},
  month = {Mar},
  publisher = {American Physical Society},
  doi = {10.1103/PhysRevLett.124.097401}
}

@article{lopezsanchez2013,
author={Lopez-Sanchez, O.
and Lembke, D.
and Kayci, M.
and Radenovic, A.
and Kis, A.},
title={{Ultrasensitive photodetectors based on monolayer ${\mathrm{MoS}}_{2}$}},
journal={Nat. Nanotechnol.},
year={2013},
month={Jul},
day={01},
volume={8},
number={7},
pages={497-501},
doi={10.1038/nnano.2013.100}
}

@book{mahan,
author = {Mahan, G. D.},
title = {{Many-Particle Physics}},
publisher = {Springer New York},
year = {2010},
doi = {10.1007/978-1-4757-5714-9},
edition   = {3rd},
}

@article{mak2013,
author={Mak, K. F.
and He, K.
and Lee, C.
and Lee, G. H.
and Hone, J.
and Heinz, T. F.
and Shan, J.},
title={{Tightly bound trions in monolayer ${\mathrm{MoS}}_{2}$}},
journal={Nat. Mater.},
year={2013},
month={Mar},
day={01},
volume={12},
number={3},
pages={207-211},
doi={10.1038/nmat3505}
}

@article{mak2016,
author={Mak, K. F.
and Shan, J.},
title={{Photonics and optoelectronics of 2D semiconductor transition metal dichalcogenides}},
journal={Nat. Photon.},
year={2016},
month={Apr},
day={01},
volume={10},
number={4},
pages={216-226},
doi={10.1038/nphoton.2015.282}
}

@article{manzeli2019,
author={Manzeli, S.
and Dumcenco, D.
and Migliato Marega, G.
and Kis, A.},
title={{Self-sensing, tunable monolayer ${\mathrm{MoS}}_{2}$ nanoelectromechanical resonators}},
journal={Nat. Commun.},
year={2019},
month={Oct},
day={23},
volume={10},
number={1},
pages={4831},
doi={10.1038/s41467-019-12795-1}
}

@article{milczewski2024,
  title = {{Momentum-dependent quasiparticle properties of the Fermi polaron from the functional renormalization group}},
  author = {von Milczewski, J. and Schmidt, R.},
  journal = {Phys. Rev. A},
  volume = {110},
  issue = {3},
  pages = {033309},
  numpages = {18},
  year = {2024},
  month = {Sep},
  publisher = {American Physical Society},
  doi = {10.1103/PhysRevA.110.033309}
}

@article{ngampruetikorn2012,
	author = {Ngampruetikorn, V. and Levinsen, J. and Parish, M. M.},
	doi = {10.1209/0295-5075/98/30005},
	journal = {EPL},
	month = {may},
	number = {3},
	pages = {30005},
	title = {Repulsive polarons in two-dimensional Fermi gases},
	volume = {98},
	year = {2012}}

@article{parish2021,
  title = {{Thermodynamic signatures of the polaron-molecule transition in a Fermi gas}},
  author = {Parish, M. M. and Adlong, H. S. and Liu, W. E. and Levinsen, J.},
  journal = {Phys. Rev. A},
  volume = {103},
  issue = {2},
  pages = {023312},
  numpages = {14},
  year = {2021},
  month = {Feb},
  publisher = {American Physical Society},
  doi = {10.1103/PhysRevA.103.023312}
}

@article{pu2021,
author = {Pu, J. and Zhang, W. and Matsuoka, H. and Kobayashi, Y. and Takaguchi, Y. and Miyata, Y. and Matsuda, K. and Miyauchi, Y. and Takenobu, T.},
title = {{Room-Temperature Chiral Light-Emitting Diode Based on Strained Monolayer Semiconductors}},
journal = {Adv. Mater.},
volume = {33},
number = {36},
pages = {2100601},
doi = {https://doi.org/10.1002/adma.202100601},
year = {2021}
}

@article{rana2020,
  title = {{Many-body theory of the optical conductivity of excitons and trions in two-dimensional materials}},
  author = {Rana, F. and Koksal, O. and Manolatou, C.},
  journal = {Phys. Rev. B},
  volume = {102},
  issue = {8},
  pages = {085304},
  numpages = {13},
  year = {2020},
  month = {Aug},
  publisher = {American Physical Society},
  doi = {10.1103/PhysRevB.102.085304}
}

@Article{ross2013,
author={Ross, J. S.
and Wu, S.
and Yu, H.
and Ghimire, N. J.
and Jones, A. M.
and Aivazian, G.
and Yan, J.
and Mandrus, D. G.
and Xiao, D.
and Yao, W.
and Xu, X.},
title={{Electrical control of neutral and charged excitons in a monolayer semiconductor}},
journal={Nat. Commun.},
year={2013},
month={Feb},
day={12},
volume={4},
number={1},
pages={1474},
doi={10.1038/ncomms2498}
}

@article{santos2013,
    author = {Santos, E. J. G. and Kaxiras, E.},
    title = {{Electrically Driven Tuning of the Dielectric Constant in ${\mathrm{MoS}}_{2}$ Layers}},
    journal = {ACS Nano},
    volume = {7},
    number = {12},
    pages = {10741-10746},
    year = {2013},
    month = {11},
    doi = {10.1021/nn403738b}
}

@article{scazza2017,
  title = {{Repulsive Fermi Polarons in a Resonant Mixture of Ultracold $^{6}\mathrm{Li}$ Atoms}},
  author = {Scazza, F. and Valtolina, G. and Massignan, P. and Recati, A. and Amico, A. and Burchianti, A. and Fort, C. and Inguscio, M. and Zaccanti, M. and Roati, G.},
  journal = {Phys. Rev. Lett.},
  volume = {118},
  issue = {8},
  pages = {083602},
  numpages = {6},
  year = {2017},
  month = {Feb},
  publisher = {American Physical Society},
  doi = {10.1103/PhysRevLett.118.083602}
}

@article{schaibley2016,
author={Schaibley, J. R.
and Yu, H.
and Clark, G.
and Rivera, P.
and Ross, J. S.
and Seyler, K. L.
and Yao, W.
and Xu, X.},
title={{Valleytronics in 2D materials}},
journal={Nat. Rev. Mat.},
year={2016},
month={Aug},
day={23},
volume={1},
number={11},
pages={16055}, 
doi={10.1038/natrevmats.2016.55}
}

@article{schmidt2012,
  title = {{Fermi polarons in Two Dimensions}},
  author = {Schmidt, R. and Enss, T. and Pietil{\"a}, V. and Demler, E.},
  journal = {Phys. Rev. A},
  volume = {85},
  issue = {2},
  pages = {021602},
  numpages = {5},
  year = {2012},
  month = {Feb},
  publisher = {American Physical Society},
  doi = {10.1103/PhysRevA.85.021602}
}

@article{shen2021,
author={Shen, P.-C.
and Su, C.
and Lin, Y.
and Chou, A.-S.
and Cheng, C.-C.
and Park, J.-H.
and Chiu, M.-H.
and Lu, A.-Y.
and Tang, H.-L.
and Tavakoli, M. M.
and Pitner, G.
and Ji, X.
and Cai, Z.
and Mao, N.
and Wang, J.
and Tung, V.
and Li, J.
and Bokor, J.
and Zettl, A.
and Wu, C.-I
and Palacios, T.
and Li, L.-J.
and Kong, J.},
title={{Ultralow contact resistance between semimetal and monolayer semiconductors}},
journal={Nature},
year={2021},
month={May},
day={01},
volume={593},
number={7858},
pages={211-217},
doi={10.1038/s41586-021-03472-9}
}

@article{shi2013,
  title = {{Quasiparticle band structures and optical properties of strained monolayer ${\mathrm{MoS}}_{2}$ and ${\mathrm{WS}}_{2}$}},
  author = {Shi, H. and Pan, H. and Zhang, Y.-W. and Yakobson, B. I.},
  journal = {Phys. Rev. B},
  volume = {87},
  issue = {15},
  pages = {155304},
  numpages = {8},
  year = {2013},
  month = {Apr},
  publisher = {American Physical Society},
  doi = {10.1103/PhysRevB.87.155304}
}

@article{sidler2016,
doi = {10.1038/nphys3949},  
year = {2017},
volume = {13},
number = {3},
pages = {255-261},
author = {Sidler, M. and Back, P. and Cotlet, O. and  Srivastava, A. and Fink, T. and Kroner, M. and Demler, E. and Imamoglu, A.},
title = {{Fermi Polaron-Polaritons in Charge-Tunable Atomically Thin Semiconductors}},
journal = {Nat. Phys.}
}

@article{smithe2018,
    author = {Smithe, K.
K. H. and English, C. D. and Suryavanshi, S. V. and Pop, E.},
    title = {{High-Field
Transport and Velocity Saturation in Synthetic
Monolayer $\mathrm{Mo}{\mathrm{S}}_{2}$}},
    journal = {Nano Lett.},
    volume = {18},
    number = {7},
    pages = {4516-4522},
    year = {2018},
    month = {06},
    doi = {10.1021/acs.nanolett.8b01692}
}

@article{smolenski2019,
  title = {{Interaction-Induced Shubnikov--de Haas Oscillations in Optical Conductivity of Monolayer ${\mathrm{MoSe}}_{2}$}},
  author = {Smole\ifmmode \acute{n}\else \'{n}\fi{}ski, T. and Cotlet, O. and Popert, A. and Back, P. and Shimazaki, Y. and Kn\"uppel, P. and Dietler, N. and Taniguchi, T. and Watanabe, K. and Kroner, M. and Imamoglu, A.},
  journal = {Phys. Rev. Lett.},
  volume = {123},
  issue = {9},
  pages = {097403},
  numpages = {6},
  year = {2019},
  month = {Aug},
  publisher = {American Physical Society},
  doi = {10.1103/PhysRevLett.123.097403}
}

@article{son2019,
  title = {{Strain Engineering of the Berry Curvature Dipole and Valley Magnetization in Monolayer ${\mathrm{MoS}}_{2}$}},
  author = {Son, J. and Kim, K.-H. and Ahn, Y. H. and Lee, H.-W. and Lee, J.},
  journal = {Phys. Rev. Lett.},
  volume = {123},
  issue = {3},
  pages = {036806},
  numpages = {6},
  year = {2019},
  month = {Jul},
  publisher = {American Physical Society},
  doi = {10.1103/PhysRevLett.123.036806}
}

@article{strasbourg2025,
    author = {Strasbourg, M.
C. and Yanev, E. S. and Parvez, S. and Stage, J. and Darlington, T. P. and Liu, S. and Holbrook, M. A. and Hone, J. C. and S., P. J. and Borys, N. J.},
    title = {{Structure-Dependent
Dynamics of Quantum Emitters in
Single-Layer $\mathrm{W}{\mathrm{Se}}_{2}$}},
    journal = {Nano Lett.},
    volume = {25},
    number = {25},
    pages = {10254-10261},
    year = {2025},
    month = {06},
    doi = {10.1021/acs.nanolett.5c02620}
}

@article{tiene2023,
  title = {{Crossover from exciton polarons to trions in doped two-dimensional semiconductors at finite temperature}},
  author = {Tiene, A. and Mulkerin, B. C. and Levinsen, J. and Parish, M. M. and Marchetti, F. M.},
  journal = {Phys. Rev. B},
  volume = {108},
  issue = {12},
  pages = {125406},
  numpages = {18},
  year = {2023},
  month = {Sep},
  publisher = {American Physical Society},
  doi = {10.1103/PhysRevB.108.125406}
}

@misc{vaz2026,
      title={{Counterdirectional Exciton and Trion Motion in Applied Electric Field}}, 
      author={D. Vaz and Y. Guan and Q. Wan and B. Bobby and A. Ramavath and B. Vargo and J. Ye and J. Beaumariage and O. Patel and K. Watanabe and T. Taniguchi and X. Feng and J. Hone and N. Youngblood and Z. Sun and D. W. Snoke},
      year={2026},
      eprint={2607.22250},
      archivePrefix={arXiv},
      url={https://arxiv.org/abs/2607.22250}, 
}

@article{vitale2018,
author = {Vitale, S. A. and Nezich, D. and Varghese, J. O. and Kim, P. and Gedik, N. and Jarillo-Herrero, P. and Xiao, D. and Rothschild, M.},
title = {{Valleytronics: Opportunities, Challenges, and Paths Forward}},
journal = {Small},
volume = {14},
number = {38},
pages = {1801483},
doi = {https://doi.org/10.1002/smll.201801483},
year = {2018}
}

@article{wagner2021,
  title = {{Nonclassical Exciton Diffusion in Monolayer ${\mathrm{WSe}}_{2}$}},
  author = {Wagner, K. and Zipfel, J. and Rosati, R. and Wietek, E. and Ziegler, J. D. and Brem, S. and Perea-Caus\'{\i}n, R. and Taniguchi, T. and Watanabe, K. and Glazov, M. M. and Malic, E. and Chernikov, A.},
  journal = {Phys. Rev. Lett.},
  volume = {127},
  issue = {7},
  pages = {076801},
  numpages = {8},
  year = {2021},
  month = {Aug},
  publisher = {American Physical Society},
  doi = {10.1103/PhysRevLett.127.076801}
}

@article{wang2012,
    author = {Wang, H. and Yu, L. and Lee, Y.-H. and S., Y. and H., A. and C., M.
L. and Li, L.-J. and Dubey, M. and Kong, J. and Palacios, T.},
    title = {{Integrated Circuits Based
on Bilayer ${\mathrm{MoS}}_{2}$ Transistors}},
    journal = {Nano Lett.},
    volume = {12},
    number = {9},
    pages = {4674-4680},
    year = {2012},
    month = {08},
    doi = {10.1021/nl302015v}
}

@article{wang2020,
  title = {{Observation of Quantized Exciton Energies in Monolayer ${\mathrm{WSe}}_{2}$ under a Strong Magnetic Field}},
  author = {Wang, T. and Li, Z. and Lu, Z. and Li, Y. and Miao, S. and Lian, Z. and Meng, Y. and Blei, M. and Taniguchi, T. and Watanabe, K. and Tongay, S. and Yao, W. and Smirnov, D. and Zhang, C. and Shi, S.-F.},
  journal = {Phys. Rev. X},
  volume = {10},
  issue = {2},
  pages = {021024},
  numpages = {9},
  year = {2020},
  month = {Apr},
  publisher = {American Physical Society},
  doi = {10.1103/PhysRevX.10.021024}
}

@article{wang2022,
author = {Wang, Y. and Iyikanat, F. and Rostami, H. and Bai, X. and Hu, X. and Das, S. and Dai, Y. and Du, L. and Zhang, Y. and Li, S. and Lipsanen, H. and Garc\'ia de Abajo, F. J. and Sun, Z.},
title = {{Probing Electronic States in Monolayer Semiconductors through Static and Transient Third-Harmonic Spectroscopies}},
journal = {Adv. Mater.},
volume = {34},
number = {3},
pages = {2107104},
doi = {https://doi.org/10.1002/adma.202107104},
year = {2022}
}

@Article{ye2016,
author={Ye, Y.
and Xiao, J.
and Wang, H.
and Ye, Z.
and Zhu, H.
and Zhao, M.
and Wang, Y.
and Zhao, J.
and Yin, X.
and Zhang, X.},
title={{Electrical generation and control of the valley carriers in a monolayer transition metal dichalcogenide}},
journal={Nat. Nanotechnol.},
year={2016},
month={Jul},
day={01},
volume={11},
number={7},
pages={598-602},
doi={10.1038/nnano.2016.49}
}

@article{zeng2018,
author = {Zeng, Q. and Liu, Z.},
title = {{Novel Optoelectronic Devices: Transition-Metal-Dichalcogenide-Based 2D Heterostructures}},
journal = {Adv. Electron. Mater.},
volume = {4},
number = {2},
pages = {1700335},
doi = {https://doi.org/10.1002/aelm.201700335},
year = {2018}
}

@article{zhang2014,
  title = {{Absorption of light by excitons and trions in monolayers of metal dichalcogenide $\mathrm{Mo}{\mathrm{S}}_{2}$: Experiments and theory}},
  author = {Zhang, C. and Wang, H. and Chan, W. and Manolatou, C. and Rana, F.},
  journal = {Phys. Rev. B},
  volume = {89},
  issue = {20},
  pages = {205436},
  numpages = {10},
  year = {2014},
  month = {May},
  publisher = {American Physical Society},
  doi = {10.1103/PhysRevB.89.205436},
}

@article{zhao2021,
doi = {10.1088/1361-6633/abdb98},
year = {2021},
month = {mar},
publisher = {IOP Publishing},
volume = {84},
number = {2},
pages = {026401},
author = {Zhao, S. and Li, X. and Dong, B. and Wang, H. and Wang, H. and Zhang, Y. and Han, Z. and Zhang, H.},
title = {{Valley manipulation in monolayer transition metal dichalcogenides and their hybrid systems: status and challenges}},
journal = {Rep. Prog. Phys.}
}

@article{zhu2015,
author={Zhu, B.
and Chen, X.
and Cui, X.},
title={{Exciton Binding Energy of Monolayer ${\mathrm{WS}}_{2}$}},
journal={Sci. Rep.},
year={2015},
month={Mar},
day={18},
volume={5},
number={1},
pages={9218},
doi={10.1038/srep09218}
}

@article{zipfel2020,
  title = {{Exciton diffusion in monolayer semiconductors with suppressed disorder}},
  author = {Zipfel, J. and Kulig, M. and Perea-Caus\'{\i}n, R. and Brem, S. and Ziegler, J. D. and Rosati, R. and Taniguchi, T. and Watanabe, K. and Glazov, M. M. and Malic, E. and Chernikov, A.},
  journal = {Phys. Rev. B},
  volume = {101},
  issue = {11},
  pages = {115430},
  numpages = {20},
  year = {2020},
  month = {Mar},
  publisher = {American Physical Society},
  doi = {10.1103/PhysRevB.101.115430}
}

@article{liu2025,
  author  = {Liu, C. and Jia, T. and Sun, Z. and
             Gu, Y. and Xu, F. and Watanabe, K. and
             Taniguchi, T. and Jia, J. and Wang, S. and
             Liu, X. and Li, T.},
  title   = {{Density-dependent spin susceptibility and effective mass
             in monolayer {MoSe$_2$}}},
  journal = {2D Mater.},
  volume  = {12},
  number  = {3},
  pages   = {035005},
  year    = {2025},
  doi     = {10.1088/2053-1583/adce18}
}

@misc{anthropic2026claude,
  author       = {{Anthropic}},
  title        = {Claude [large language model},
  howpublished = {\url{https://claude.ai}},
}

@article{ExchangeEffects,
  title = {Effect of fermion indistinguishability on optical absorption of doped two-dimensional semiconductors},
  author = {Tiene, A. and Levinsen, J. and Keeling, J. and Parish, M. M. and Marchetti, F. M.},
  journal = {Phys. Rev. B},
  volume = {105},
  issue = {12},
  pages = {125404},
  numpages = {19},
  year = {2022},
  month = {Mar},
  publisher = {American Physical Society},
  doi = {10.1103/PhysRevB.105.125404},
  url = {https://link.aps.org/doi/10.1103/PhysRevB.105.125404}
}

\end{document}